\documentclass[journal=jctcce,manuscript=article,layout=onecolumn]{achemso}

\usepackage{graphicx}
\usepackage{dcolumn}%
\usepackage{bm}%
\usepackage{latexsym}
\usepackage{amsmath}
\usepackage{amssymb}
\usepackage{amsfonts}
\usepackage{xcolor}
\usepackage{braket}
\usepackage{booktabs}

\def\br{\mathbf{r}}

\def\h2o{\mathrm{H}_2\mathrm{O}}

\def\icomp{\mathrm{i}}

\graphicspath{{./figs/}{./}}

\title{Fully dynamic $G3W2$ self-energy for finite systems: Formulas and benchmark}

\author{Fabien Bruneval}
\email{fabien.bruneval@cea.fr}
\affiliation{Universit\' e Paris-Saclay, CEA, Service de recherche en Corrosion et Comportement des Matériaux, SRMP, 91191 Gif-sur-Yvette, France}

\author{Arno Förster}
\email{a.t.l.foerster@vu.nl}
\affiliation{Theoretical Chemistry, Vrije Universiteit, De Boelelaan 1108, 1081 HZ Amsterdam, The Netherlands}

\date{\today}

\begin{document}

\begin{abstract}
Over the years, Hedin's $GW$ self-energy has been proven to be a rather accurate and simple
approximation to evaluate electronic quasiparticle energies in solids and in molecules.
Attempts to improve over the simple $GW$ approximation, the so-called vertex corrections,
have been constantly proposed in the literature.
Here, we derive, analyze, and benchmark the complete second-order term in the screened Coulomb interaction $W$
for finite systems.
This self-energy named $G3W2$
contains all the possible time orderings
that combine 3 Green's functions $G$ and 2 dynamic $W$.
We present the analytic formula and its imaginary frequency counterpart, the
latter allowing us to treat larger molecules.
The accuracy of the $G3W2$ self-energy is evaluated
on well-established benchmarks (GW100, Acceptor 24 and Core 65) for valence and core quasiparticle energies. 
Its link with the simpler static approximation, named SOSEX for static screened
second-order exchange, is analyzed, which leads us to propose a more consistent approximation
named 2SOSEX. 
% Notably, both our formula for the $G3W2$ expression and our numerical results differ from the expression published in a recent study by Wang \textit{et. al.}
% [J. Chem. Theory Comput. \textbf{17}, 5140 (2021)]
In the end, we find that neither the $G3W2$ self-energy nor any of the investigated approximations to it improve over one-shot $G_0W_0$ with a good starting point.
Only quasi-particle self-consistent $GW$ HOMO energies are slightly improved by addition of the $G3W2$ self-energy correction. 
We show that this is due to the self-consistent update of the screened Coulomb interaction leading to an overall sign change of the vertex correction to the frontier quasiparticle energies.
\end{abstract}

\begin{tocentry}
\includegraphics[scale=0.3]{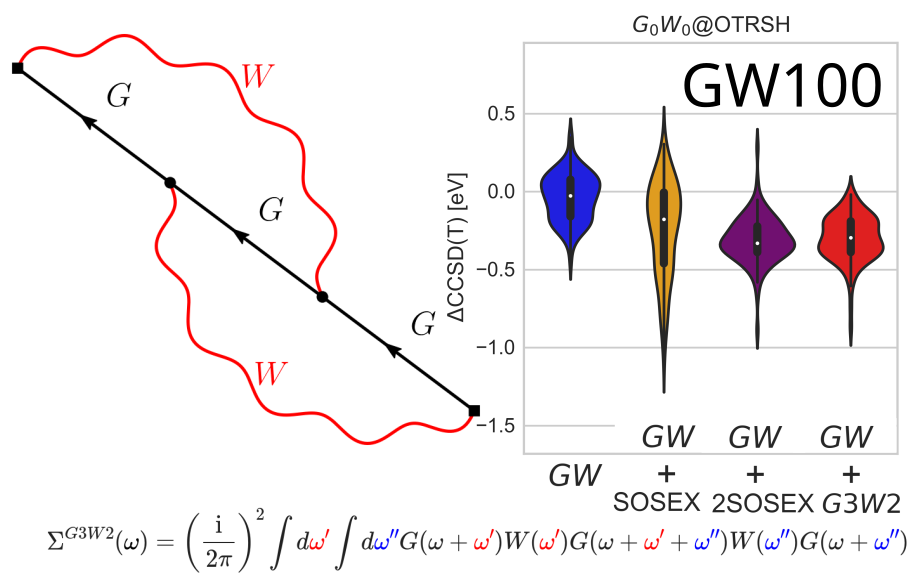}
\end{tocentry}

\maketitle

%%%%%%%%%%%%%%%%%%%%%%%%%%%%%%%%%%%%%%%%%%%%%%%%%%%%%%%%%%%%%%%%%%%%%%%%%%%
\section{Introduction}
\label{sec:intro}
%%%%%%%%%%%%%%%%%%%%%%%%%%%%%%%%%%%%%%%%%%%%%%%%%%%%%%%%%%%%%%%%%%%%%%%%%%%

Hedin's $GW$ approximation \cite{hedin_pr1965} has become a wide-spread method to
evaluate the electronic quasiparticle energies \cite{reining_wires2018,golze_fchem2019}.
While first introduced for extended systems
\cite{hedin_pr1965,lundqvist_pkm1967,strinati_prl1980,hybertsen_prl1985,godby_prl1986},
it has recently permeated to molecular systems
\cite{shirley_prb1993,grossman_prl2001,rostgaard_prb2010,blase_prb2011,bruneval_jcp2012,ren_njp2012,sharifzadeh_epjb2012,korzdorfer_prb2012,bruneval_jctc2013,vansetten_jctc2013,koval_prb2014,govoni_jctc2015,vansetten_jctc2015,knight_jctc2016,kawahara_prb2016,blase_jcp2016,hesselmann_pra2017,maggio_jctc2017b,Vlcek2017,Vlcek2018, lange_jctc2018,wilhelm_jcpl2018,veril_jctc2018,foerster_jctc2020, Bintrim2021},
and even more recently to the core state binding energy \cite{golze_jctc2018,vansetten_jctc2018,golze_jpcl2020}.
The success of the $GW$ approximation owes much to its cost-effectiveness:
$GW$ is both rather simple and reasonably accurate.

However, almost as old as $GW$ itself, there have been attempts to go beyond the mere $GW$ self-energy.
In the framework of the many-body equations as formulated by Hedin \cite{hedin_pr1965,hedin_chapter1970},
these corrections are coined ``vertex correction'', since they necessarily involve
the complicated 3-point vertex function $\tilde \Gamma$.
The vertex corrections appear in two distinct locations in the equations,
firstly as a correction to the polarizability and secondly as a correction to the self-energy itself.
The first kind of vertex corrections are less computationally demanding
and have been tried first
\cite{northrup_prl1987,shirley_prb1993,bruneval_prl2005a,shishkin_prl2007}.
But a recent work \cite{lewis_jctc2019} indicates the vertex corrections in the polarizability
systematically only have a mild effect,
even though there exists a series of studies that include both vertices at once.
\cite{delsole_prb1994,maggio_jctc2017a,Vlcek2019,Mejuto-Zaera2021}

\begin{figure}
\includegraphics[width=0.99\columnwidth]{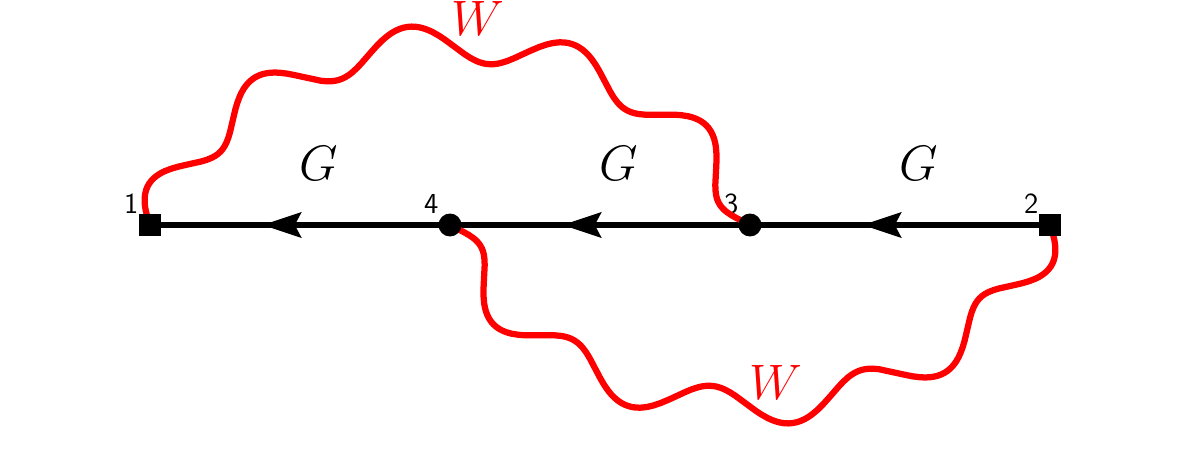}
\caption{Dynamical $G3W2$ Goldstone-Feynman diagram.
Red wiggly lines represent the screened Coulomb interactions $W$,
straight black lines with an arrow the Green's functions $G$.
Entry and exit points of the self-energy are marked with a square.}
\label{fig:diag_g3w2}
\end{figure}

Here, we focus on the latter type of vertex corrections: those that modify the electronic self-energy itself.
The polarizability is kept at the simple random-phase approximation level.
Quite the opposite, for the self-energy the intrinsic 3-point nature of the vertex function cannot be avoided and
the practical derivations and calculations have to cope with this additional complexity.
Here we focus in particular on the second-order diagram that we name $G3W2$
following Kutepov \cite{kutepov_jpcm2021}
and whose Feynman-Goldstone representation is given in Figure~\ref{fig:diag_g3w2}.
Gradual progress has been made towards the complete inclusion of the second-order term.
Most of the complexity arises from the screened Coulomb interaction $W(\omega)$ that is 
a dynamic quantity as opposed to the bare Coulomb interaction $v$.
Earlier simpler approximations such as the second-order exchange (SOX)
and the second-order screened exchange (SOSEX) exist.
 SOX is obtained when the two $W$ are replaced by $v$.
This approximation has been known for long \cite{szabo_book}
but its performance is very poor \cite{korzdorfer_prb2012,bruneval_fchem2021}.
The next step, SOSEX, consists of one dynamic $W(\omega)$ and one static $v$
\cite{ren_prb2015}.
SOSEX performance is somewhat better than SOX \cite{bruneval_fchem2021}.
The complete calculation of the dynamic $G3W2$ term has been recently claimed
by Wang, Rinke, and Ren \cite{wang_jctc2021}.
Note that statically screened approximations to $G3W2$ using $W(\omega=0)$ instead of
the dynamic $W(\omega)$ have also been proposed in the literature
\cite{gruneis_prl2014} with interesting performance.\cite{foerster_prb2022, Forster2023a}
There are clues from the homogeneous electron gas model
that the $G3W2$ diagram contains undesired features for 
extended systems \cite{stefanucci_prb2014,pavlyukh_prl2016}.
However an evaluation for molecular systems is missing as of today.

In this article, we concentrate on the fully dynamic $G3W2$ self-energy.
We derive its analytic formula as well as its complex frequency numerical quadrature.
Our expressions depart from the one reported in Ref.~\citenum{wang_jctc2021}.
A double frequency integral has to be performed in our equations,
which makes the practical calculations very challenging.
We analyze the different contributions to the $G3W2$ self-energy induced by the possible different
time-orderings.
Consistent numerical results are obtained with different expressions and two independent codes,
MOLGW \cite{bruneval_cpc2016} and BAND.\cite{TeVelde1991, Philipsen2022}
Thanks to our analytic expressions for both $G3W2$ and SOSEX, we are able to accurately evaluate
them for the core states.
We then test the performance of $G3W2$ and its approximations (SOX, SOSEX) on well-established
benchmarks (GW100 \cite{vansetten_jctc2015}, Acceptor24\cite{richard_jctc2016}, and 
a subset of CORE65\cite{golze_jpcl2020}).

%%%%%%%%%%%%%%%%%%%%%%%%%%%%%%%%%%%%%%%%%%%%%%%%%%%%%%%%%%%%%%%%%%%%%%%%%%%
\section{$G3W2$ self-energy derivation}
\label{sec:theory}
%%%%%%%%%%%%%%%%%%%%%%%%%%%%%%%%%%%%%%%%%%%%%%%%%%%%%%%%%%%%%%%%%%%%%%%%%%%

\subsection{$G3W2$ self-energy obtained from Hedin's equations}

There are many equivalent ways to obtain the analytic expression of the self-energy
diagram drawn in Fig.~\ref{fig:diag_g3w2}.
We follow here the functional path as exposed by Hedin in his seminal paper\cite{hedin_pr1965}.
More specifically, we closely stick to Strinati's review notations\cite{strinati_rnc1988},
except for the exchange-correlation self-energy that we denote $\Sigma$ instead of $M$.

The exact electronic self-energy $\Sigma$ reads
\begin{equation}
\label{eq:exact_sigma}
\Sigma(1,2) =  \icomp
   \int d3 d4 G(1, 4) W(1^+, 3) \tilde \Gamma(4,2;3)  ,
\end{equation}
where the numbers, such as 1, conventionally represent a combined index over space and time ($\br_1$,$t_1$).
$G$ is the time-ordered Green's function,
$W$ is the dynamically screened Coulomb interaction, and
$\tilde \Gamma(4,2;3)$ is the irreducible 3-point vertex function.
These three functions have been already mentioned in the introduction.
According to Strinati \cite{strinati_rnc1988},
the irreducible vertex has a closed expression:
\begin{equation}
\label{eq:gamma}
  \tilde \Gamma(4,2;3) = \delta(4,2) \delta(4,3)
   +  \int d5678 \frac{\delta \Sigma(4,2)}{\delta G(5,6)}
          G(5,7) G(8,6)   \tilde \Gamma(7,8;3)   .
\end{equation}

Obviously, Eq.~(\ref{eq:gamma}) is tremendously complex involving 3- and 4-point functions.
The usual $GW$ approximation is obtained when Eq.~(\ref{eq:gamma}) is truncated to the mere
$\delta$-functions (first term in the right-hand side).
This approximation introduced in Eq.~(\ref{eq:exact_sigma}) indeed yields
\begin{equation}
\label{eq:gw}
\Sigma(1,2) =  \icomp G(1, 2) W(1^+, 2) .
\end{equation}

Let us approximate $\tilde \Gamma(4,2;3)$ in a slightly more sophisticated way.
In Eq.~(\ref{eq:gamma}), consider that the irreducible vertex in the right-hand side
is simply
\begin{subequations}
\begin{equation}
\label{eq:tilde_approx}
 \tilde \Gamma(7,8;3) \approx \delta(7,8) \delta(7,3) .
\end{equation}
In addition, use the $GW$ self-energy in the derivative of $\Sigma$ with respect to $G$
in the same Eq.~(\ref{eq:gamma}), which gives
\begin{equation}
\label{eq:vertexApproximation}
  \frac{\delta \Sigma(4,2)}{\delta G(5,6)} \approx \icomp W(4,2) \delta(4,5) \delta(2,6)  ,
\end{equation}
which amounts to neglecting the derivative of $W$ with respect to $G$.
\end{subequations}

With these two simplifications, Eq.~(\ref{eq:gamma}) reduces to
\begin{equation}
\label{eq:gamma_simple}
 \tilde \Gamma(4,2;3) \approx
    \delta(4,2) \delta(4,3) 
       +    \icomp W(4,2) G(4,3) G(3,2)  ,
\end{equation}
which is a lot less complex than the original expression.

Plugging Eq.~(\ref{eq:gamma_simple}) into the expression of the exact self-energy reported in Eq.~(\ref{eq:exact_sigma}),
we obtain our desired expression:
\begin{equation}
\label{eq:gw_g3w2}
\Sigma(1,2) =  \icomp G(1,2) W(1^+,2) \\
    + \icomp^2  \int d3 d4 G(1, 4) W(1^+, 3) G(4,3) W(4,2) G(3,2)  .
\end{equation}

The self-energy in Eq.~(\ref{eq:gw_g3w2}) consists of two terms: the usual $GW$ approximation and
an additional correction (in other words a ``vertex correction'') that we name $G3W2$.
The diagrammatic representation of $G3W2$ precisely matches the diagram of Figure~\ref{fig:diag_g3w2}.
Note that the $G3W2$ expression has a double time integral, which still make it quite a challenge for
numerical applications.

\subsection{$G3W2$, 2SOSEX, SOSEX, and SOX}

It is insightful (and convenient for numerics as we will see later) to split the
dynamic $G3W2$ self-energy correction into static and dynamic parts thanks to the introduction
of the intermediate bare Coulomb interaction $v$.

Dropping the space and time indices for conciseness, one can decompose the $G3W2$ self-energy
into 4 terms:
\begin{eqnarray}
\label{eq:g3w2_split}
\Sigma^{G3W2} &=& \icomp^2  \int G W G W G  \nonumber \\
 &=&
\icomp^2  \int G v G v G \nonumber \\
& & + \icomp^2  \int G \left[W-v \right] G v G \nonumber \\
& & + \icomp^2  \int G v G \left[W-v \right] G  \nonumber \\
& & + \icomp^2  \int G \left[W-v \right] G \left[ W-v \right] G   .
\end{eqnarray}

The first term is the well-known SOX term \cite{szabo_book}.
The sum of the two first terms form the SOSEX self-energy introduced by Ren and coworkers \cite{ren_prb2015}.

The second and the third terms are equal since they are completely symmetric through the exchange of integrated indices.
We hence propose to name 
``2SOSEX'' the sum of the 3 first terms in Eq.~(\ref{eq:g3w2_split}).
Of course, the last term that contains a genuine time-dependence in both interactions
is the most computationally involved.

From the present perspective, there is no specific reason to use SOSEX rather than 2SOSEX.
We will study in section~\ref{sec:bench} the actual performance of each truncation strategy.

\subsection{Expression in real frequency domain}

For practical implementation and understanding, it is very convenient
to work in the frequency domain.
As the first terms in Eq.~(\ref{eq:g3w2_split}) are already known,
we now focus on the last, fully-dynamic one.

In the following we introduce $W_p$ for the polarizable part of $W$ as 
\begin{equation}
 W_p(1,2) = W(1,2) - v(1,2) \delta(t_1-t_2)  .
\end{equation}

In many-body perturbation theory at equilibrium, the time-dependencies of $\Sigma$, $G$, and $W$
are only a time differences: e.g. $G(1,2) = G(\br_1,\br_2, t_1-t_2)$ \cite{szabo_book}.
Omitting space variables for conciseness, 
the last term in Eq.~(\ref{eq:g3w2_split}) reads
\begin{equation}
 \label{eq:g3w2_times}
 \Sigma^{G3{W_p}2}(t_1-t_2) = \icomp^2 \int dt_3 dt_4 G(t_1 -t_4) W_p(t_3-t_1) G(t_4-t_3) W_p(t_2-t_4) G(t_3-t_2)  .
\end{equation}
We have used the symmetries $W_p(1,3) = W_p(3,1)$ and $W_p(4,2) = W_p(2,4)$ for convenience \cite{strinati_rnc1988}.

Then we Fourier-transform of all these functions.
The detailed steps that lead to the final expression are given in appendix~\ref{app:realfreqs}.
Finally, the fully-dynamic term $\Sigma^{G3{W_p}2}$ in the frequency domain reads
\begin{equation}
  \label{eq:g3wp2}
  \Sigma^{G3{W_p}2}(\omega) =
    \left( \frac{\icomp}{2\pi} \right)^2
     \int d\omega^\prime \int d\omega^{\prime\prime}
        G(\omega+\omega^\prime)
        W_p(\omega^\prime) 
        G(\omega+\omega^\prime+\omega^{\prime\prime})
        W_p(\omega^{\prime\prime})
        G(\omega+\omega^{\prime\prime})   .
\end{equation}

The expression with times had two integrals and it still contains two integrals in
the frequency domain.

%Now using
%\begin{multline}
%\label{eq:g3w2_real}
% \Sigma^{G3W2}_{pq}(\omega)
%   =  \left( \frac{\icomp}{2\pi} \right)^2
%    \int_{-\infty}^{+\infty} d\omega^\prime \int_{-\infty}^{+\infty} d\omega^{\prime\prime}
%    \sum_u 
%    \sum_v
%    \sum_w  \\
%\frac{( w v | W_p(\omega^\prime) | p u )
%( q w | W_p(\omega^{\prime\prime}) | u v )}
%{(\omega + \omega^\prime - \epsilon_u)
%(\omega +\omega^\prime + \omega^{\prime\prime} - \epsilon_v)
%(\omega +\omega^{\prime\prime} - \epsilon_w)}  .
%\end{multline}

\subsection{Analytic formula}

In this section, we derive a fully analytic formula for Eq.~(\ref{eq:g3wp2}).
This is not only useful for a deeper understanding, but also it is in practice, since
some implementations, such as MOLGW \cite{bruneval_cpc2016} or Turbomole, \cite{vansetten_jctc2013}
indeed use this strategy.

When starting from a mean-field Kohn-Sham or generalized Kohn-Sham self-consistent solution
with (real-valued) eigenstates $\varphi_p(\br)$ and eigenvalues $\epsilon_p$,
the corresponding Green's function $G(\omega)$ reads
\begin{equation}
  \label{eq:g0}
  G(\br,\br',\omega) =
    \sum_i^\mathrm{occ}
     \frac{\varphi_i(\br)   \varphi_i(\br^\prime)}
     {\omega - \epsilon_i -\icomp \eta }
+ \sum_a^\mathrm{virt}
     \frac{\varphi_a(\br)   \varphi_a(\br^\prime)}
     {\omega - \epsilon_a +\icomp \eta } .
\end{equation}
We use the usual convention where the indices $a, b, c$ run over virtual molecular orbitals (MO),
$i, j, k$ over occupied MO, and $p, q, u, v, w$ over all the MO.
$\eta$ symbol stands for a small positive real number that enforces the proper location of the poles
in the complex plane: above the real axis for occupied states ($\varepsilon_i < \mu$)
and below the real axis for virtual states ($\varepsilon_a > \mu$).
Note that we assume real wavefunctions without loss of generality for finite systems.

The polarizable part of $W$ also can be written analytically
provided that the Casida-like equation has been solved for the 
random-phase approximation \cite{tiago_prb2006,vansetten_jctc2013,bruneval_cpc2016}.
$W_p(\omega)$ is a sum of pairs of poles $\pm \Omega_s$,
which account for resonant and antiresonant neutral
excitations \cite{bruneval_cpc2016}:
\begin{equation}
\label{eq:wspec}
  W_p(\br,\br^\prime,\omega) =
    \sum_s w_s(\br) w_s(\br^\prime) 
       \left[ \frac{1}{\omega - \Omega_s + \icomp \eta}
             -\frac{1}{\omega + \Omega_s - \icomp \eta}
       \right]  .
\end{equation}
Here and in the following the pole index $s$ (and later also $t$) runs over the pairs of poles.
Again, the small real positive $\eta$ enforces the correct location of the poles for
a time-ordered function.
$\Omega_s$ are defined as positive energies as obtained from the diagonalization of the Casida-like
equation. The resonant excitation count amounts to $N_o \times N_v$ in a finite basis set,
where $N_o$ is the number of occupied MO and $N_v$ the number of virtual MO.

With these two expressions for the non-interacting $G$ and for $W_p$ in Eqs.~(\ref{eq:g0}) and (\ref{eq:wspec}), there is no obstacle against the complete derivation of $G3W2$
as expressed in Eq.~(\ref{eq:g3wp2}).
One can apply the complex analysis tools (contour closing and residue theorem) to each of the
$2^5=32$ cases for occupied or virtual MO in $G$ and resonant or antiresonant poles in $W_p$.

Fortunately, 14 terms are strictly zero becase of the poles' location.
The 18 remaining terms have been calculated manually.
A calculation for the term involving 3 occupied MO and 2 anti-resonant poles in $W_p$
is given as an example in Appendix~\ref{app:ooo}.

Introducing $w_s^{pq} = \int d\br \varphi_p(\br)\varphi_q(\br) w_s(\br)$,
we present the sum of the 18 terms that has been split according to
the occupation of the 3 Green's functions:

\begin{subequations}

\begin{equation}
\label{eq:ooo}
\Sigma_{pq}^{ooo}(\omega)
=
\sum_{t s}
\sum_{i j k } 
\frac{w_t^{pi} \cdot w_t^{jk} \cdot w_s^{qk} \cdot w_s^{ij}}
{(\omega  - \epsilon_i + \Omega_t - 2 \icomp \eta) \cdot (\omega - \epsilon_j + \Omega_t + \Omega_s  - 3 \icomp \eta )
\cdot (\omega - \epsilon_k + \Omega_s  - 2\icomp \eta)}
\end{equation}

\begin{equation}
\label{eq:vvv}
\Sigma_{pq}^{vvv}(\omega)
=
\sum_{t s}
\sum_{a b c } 
\frac{w_t^{pa} \cdot w_t^{bc} \cdot w_s^{qc} \cdot w_s^{ab}}
{(\omega  - \epsilon_a - \Omega_t + 2 \icomp \eta) \cdot (\omega - \epsilon_b - \Omega_t - \Omega_s  + 3 \icomp \eta )
\cdot (\omega - \epsilon_c - \Omega_s  + 2\icomp \eta)}
\end{equation}

\begin{multline}
\Sigma_{pq}^{voo+oov}(\omega)
=
\sum_{t s}
\sum_{a j k } 
\frac{w_t^{pa} \cdot w_t^{jk} \cdot w_s^{qk} \cdot w_s^{aj}}
{(\omega - \epsilon_k + \Omega_s  - 2 \icomp \eta)
\cdot (\Omega_s + \epsilon_a - \epsilon_j  - 3 \icomp \eta )
} \\
\times
\left[
\frac{2}{\omega - \epsilon_a - \Omega_t + 2\icomp \eta}
- \frac{2}{\omega -\epsilon_j + \Omega_s + \Omega_t  - 3\icomp \eta}
\right]   .
\end{multline}

\begin{multline}
\Sigma_{pq}^{ovv+vvo}(\omega)
=
\sum_{t s}
\sum_{i b c } 
\frac{w_t^{pi} \cdot w_t^{bc} \cdot w_s^{qc} \cdot w_s^{ib}}
{(\omega - \epsilon_c - \Omega_s  + 2 \icomp \eta)
\cdot (\Omega_s -\epsilon_i + \epsilon_b  - 3 \icomp \eta )
} \\
\times
\left[
- \frac{2}{\omega - \epsilon_i  + \Omega_t - 2\icomp \eta}
+ \frac{2}{\omega -\epsilon_b - \Omega_s - \Omega_t  + 3\icomp \eta}
\right]
\end{multline}

\begin{multline}
\Sigma_{pq}^{ovo}(\omega)
=
\sum_{t s}
\sum_{i b k} 
\frac{w_t^{pi} \cdot w_t^{bk} \cdot w_s^{qk} \cdot w_s^{ib}}
{\omega - \epsilon_i - \epsilon_k + \epsilon_b - 3 \icomp \eta} \\
\times
\left[
\frac{2 \epsilon_b - \epsilon_i - \epsilon_k + \Omega_t + \Omega_s}
{(\omega - \epsilon_b  - \Omega_t - \Omega_s + 3\icomp \eta)
\cdot
( \Omega_s  + \epsilon_b - \epsilon_i  - 3 \icomp \eta)
\cdot
(  \Omega_t + \epsilon_b - \epsilon_k  - 3 \icomp \eta)}
\right.\\
  - \frac{2}{(\Omega_t + \epsilon_b - \epsilon_k  - 3\icomp \eta)
 \cdot 
 ( \omega - \epsilon_k + \Omega_s - 2\icomp\eta ) } \\ 
 \left.
- \frac{1}{(\omega -\epsilon_i + \Omega_t  - 2\icomp \eta)
 \cdot 
 ( \omega - \epsilon_k + \Omega_s - 2\icomp\eta ) }
\right]
\end{multline}

\begin{multline}
\label{eq:vov}
\Sigma_{pq}^{vov}(\omega)
=
\sum_{t s}
\sum_{a j c} 
\frac{w_t^{pa} \cdot w_t^{jc} \cdot w_s^{qc} \cdot w_s^{aj}}
{\omega - \epsilon_a - \epsilon_c + \epsilon_j + 3 \icomp \eta} \\
\times
\left[
\frac{2 \epsilon_j - \epsilon_a - \epsilon_c - \Omega_t -\Omega_s}
{(\omega - \epsilon_j  + \Omega_t + \Omega_s - 3\icomp \eta)
\cdot
( \Omega_s  - \epsilon_j + \epsilon_a  - 3 \icomp \eta)
\cdot
(  \Omega_t - \epsilon_j + \epsilon_c- 3 \icomp \eta)}
\right.\\
 + \frac{2}{(\Omega_t - \epsilon_j + \epsilon_c  - 3\icomp \eta)
 \cdot 
 ( \omega - \epsilon_c - \Omega_s + 2 \icomp\eta ) } \\ 
 \left.
- \frac{1}{(\omega - \epsilon_a - \Omega_t  + 2\icomp \eta)
 \cdot 
 ( \omega - \epsilon_c - \Omega_s    + 2 \icomp\eta ) }
\right]
\end{multline}
\end{subequations}

In the previous equations, we have kept explicit the counts of $\eta$ for debugging.
The locations of the poles appear as consistent for 
a fermionic time-ordered function, with poles above the real axis for 
$(\epsilon_i - \Omega_s)$ and for $(\epsilon_i + \epsilon_k - \epsilon_b)$
and 
poles below the real axis for $(\epsilon_a + \Omega_s)$
and for $(\epsilon_a - \epsilon_c + \epsilon_j)$.
Remember that $\Omega_s$ are positive by definition.

It should be noticed that terms with 3 occupied MO or 3 virtual MO are present.
This type of terms are absent in the static SOSEX \cite{ren_prb2015}
and its dynamic extension proposed in Ref.~\citenum{wang_jctc2021}.

The computational scaling of Eqs.~(\ref{eq:ooo}-\ref{eq:vov}) is as high as $N^7$,
since each sum over states gives $N$ and each sum over excitations yields $N^2$.

\subsection{Imaginary frequency expression}

For practical calculations, it is desirable to perform the frequency integrals thanks to a quadrature
instead.
As the poles of both $G$ and $W_p$ lie in the vicinity of the real axis,
it is customary in this field to analytically continue the function to the complete complex plane
and perform the quadrature on the imaginary axis.
More precisely, the imaginary axis should separate the occupied and virtual states.
This is achieved by selecting an origin $\mu$ that lies in the HOMO-LUMO gap.

Sometimes authors prefer to shift the energies $\epsilon_p$ so that $\mu$ is set zero
\cite{rieger_cpc1998,ren_prb2015,liu_prb2016}.
While this procedure is perfectly valid, it may mask some of analytic properties of the formulas.
In the following, we write the imaginary frequency expression of $G3W2$
with explicit $\mu$ dependencies:
\begin{multline}
\label{eq:g3w2_imag}
 \Sigma^{G3W_p2}_{pq}(\mu + \icomp\omega)
   =  \left( \frac{-1}{2\pi} \right)^2
    \int_{-\infty}^{+\infty} d\omega^\prime \int_{-\infty}^{+\infty} d\omega^{\prime\prime}
    \sum_u 
    \sum_v
    \sum_w  \\
\frac{( w v | W_p(\icomp\omega^\prime) | p u )
( q w | W_p(\icomp\omega^{\prime\prime}) | u v )}
{(\mu + \icomp\omega + \icomp\omega^\prime - \epsilon_u)
(\mu + \icomp\omega +\icomp\omega^\prime +
\icomp\omega^{\prime\prime} - \epsilon_v)
(\mu + \icomp\omega + \icomp\omega^{\prime\prime} - \epsilon_w)}  .
\end{multline}

This equation is the imaginary axis counterpart of the real axis formula:
the $\icomp$ prefactor is replaced by (-1) and the frequencies $\omega$, $\omega^\prime$,
$\omega^{\prime\prime}$ always come with a $\icomp$ factor. 
Note that the chemist notation is used here:
\begin{equation}
( w v | W_p(\icomp\omega^\prime) | p u ) = \int d\br \int d\br'
   \varphi_w(\br) \varphi_v(\br) W_p(\br,\br',\icomp\omega^\prime)
   \varphi_p(\br) \varphi_u(\br) .
\end{equation}
The derivation of Eq.~\eqref{eq:g3w2_imag} is sketched in appendix~\ref{app:imagfrequencies}.
If one splits each of the sums in Eq.~\eqref{eq:g3w2_imag} into individual summations over occupied and virtual states, one can see that Eq.~\eqref{eq:g3w2_imag} splits into 8 terms. These correspond to the 8 possible time-orderings of the three Green's functions in the $G3W2$ contribution to the self-energy and can be matched with the terms Eqs.~(\ref{eq:ooo}-\ref{eq:vov}).

Equation~(\ref{eq:g3w2_imag}) can be evaluated with a double quadrature on $\omega^\prime$
and $\omega^{\prime\prime}$.
Assuming that the grid design does not change with the system size, Eq.~(\ref{eq:g3w2_imag})
presents a $N^5$ scaling with a larger prefactor, since all sums run over occupied and virtual states. 
Notably, Eq.~\eqref{eq:g3w2_imag} is different than the one in Ref.~\citenum{wang_jctc2021}, which only contains a single imaginary frequency integral. 

In the end, we obtain the self-energy for selected frequencies $(\mu + \icomp \omega)$
that needs to be analytically continued to the real frequency axis \cite{rojas_prl1995,lebegue_prb2003}.
This last mathematical operation is reliable only in the vicinity of $\mu$.
Core states cannot be reliably treated with this approach \cite{golze_jctc2018}.

From the imaginary frequency expression in Eq.~\eqref{eq:g3w2_imag}, various approximations to the full $G3W2$ self-energy are obtained.
If one replaces one of the $W_p$ by $v$ one obtains the SOSEX self-energy expression of Ren and coworkers.\cite{ren_prb2015} Choosing $W(\icomp \omega^{\prime \prime}) = v$ allows us to perform the integral over $\omega^{\prime \prime}$ in Eq.~\eqref{eq:g3w2_imag} analytically,
\begin{equation}
\begin{aligned}
\label{eq:sosex_imag}
\Sigma^\mathrm{SOSEX}_{pq}(\mu + \icomp\omega) 
   =  & - \frac{1}{2\pi}
    \int d\omega^\prime
    \sum_u 
    \sum_v
    \sum_w 
(f_v - f_w)
\frac{
( w v | W_p(\icomp\omega^\prime) | p u )
( q w | u v )}
{(\mu + \icomp\omega + \icomp\omega^\prime - \epsilon_u)(\icomp\omega^{\prime} + \epsilon_v - \epsilon_w)} \\ 
= & 
- \frac{1}{2\pi}
    \int d\omega^\prime
    \sum_u 
    \sum_v
    \sum_w 
(f_v - f_w)
\frac{
( u v | q w  )
( p u | W_p(\icomp\omega^\prime) | v w )
}
{(\icomp\omega^{\prime} + \epsilon_v - \epsilon_w)
(\mu + \icomp\omega + \icomp\omega^\prime - \epsilon_u)} \;,
\end{aligned}
\end{equation}
where the $f_v$ denote occupation numbers, i.e. $f_v = 1$ if $v$ denotes an occupied states, and $f_v = 0$ if $v$ denotes a virtual state. The factor $(f_u - f_v)$ effectively sets half of the 8 terms in Eq.~\eqref{eq:g3w2_imag} to zero. 
If we instead replace $W(\icomp \omega^{\prime})$ by $v$, we get 
\begin{equation}
\Sigma^\mathrm{SOSEX}_{pq}(\mu + \icomp\omega)
   =  - \frac{1}{2\pi}
    \int d\omega^{\prime\prime}
    \sum_u 
    \sum_v
    \sum_w 
(f_u - f_v)\frac{( w v | p u )
( q w | W_p(\icomp\omega^{\prime\prime}) | u v )}
{(\icomp\omega^{\prime\prime} + \epsilon_u - \epsilon_v)
(\mu + \icomp\omega +\icomp\omega^{\prime\prime} - \epsilon_w)}  \;.
\end{equation}
One can see that both expressions are equivalent upon appropriate change of variables.

Replacing both $W_p$ by $v$  ($W_p(\icomp\omega^{\prime}) \rightarrow v$) and distinguishing explicitly between occupied and virtual states, one obtains the SOX self-energy
\begin{equation}
\label{eq:sox_imag}
\Sigma^\mathrm{SOX}_{pq}(\mu + \icomp\omega)
   =  
 -  \sum^\mathrm{occ}_{i,j} 
    \sum^\mathrm{virt}_a
\frac{(p i | j a)
( q j | i a )}
{\mu + \icomp\omega - \epsilon_i + \epsilon_a - \epsilon_j} 
-   \sum^\mathrm{virt}_{a,b} 
    \sum^\mathrm{occ}_i
    \frac{(p a | b i )
( q b | a i )}
{\mu + \icomp \omega - \epsilon_a + \epsilon_i - \epsilon_b} \;,
\end{equation}
in which one might replace $v$ by $W_p(\icomp \omega = 0)$ to obtain the statically screened $G3W2$ self-energy.\cite{gruneis_prl2014, foerster_prb2022} Due to the static nature of the interactions in Eq.~\eqref{eq:sox_imag}, only 2 of the 8 possible time-orderings in Eq.~\eqref{eq:g3w2_imag} have a non-zero contribution to the self-energy.

%%%%%%%%%%%%%%%%%%%%%%%%%%%%%%%%%%%%%%%%%%%%%%%%%%%%%%%%%%%%%%%%%%%%%%%%%%%
\section{Results for molecules}
\label{sec:bench}
%%%%%%%%%%%%%%%%%%%%%%%%%%%%%%%%%%%%%%%%%%%%%%%%%%%%%%%%%%%%%%%%%%%%%%%%%%%

\subsection{Implementations in MOLGW and BAND}

The formulas presented in the previous section have been implemented in two
independent quantum chemistry codes, MOLGW\cite{bruneval_cpc2016} and BAND\cite{TeVelde1991, Philipsen2022}.

In MOLGW, we have implemented both the fully analytic expression reported in Eqs.~(\ref{eq:ooo}-\ref{eq:vov})
and the imaginary frequency expression from Eq.~(\ref{eq:g3w2_imag}).
In BAND, we have focused on numerical efficiency and only the imaginary frequency technique has been implemented. 
The results in Sec.~\ref{sec:bench:basic} and \ref{sec:bench:core} have been obtained with the analytical expression in MOLGW. The majority of calculations for GW100 and ACC24 have been performed using the imaginary frequency expression in BAND. The systems containing 5th row atoms in GW100 atoms require the use of effective core potentials (ECP) for def2-TZVPP, which are not implemented in BAND. Therefore, also these calculations have been performed with MOLGW.

\subsubsection{Technical Details}

\paragraph{MOLGW}

Both implementations in MOLGW (analytic and imaginary frequency quadrature)
can use frozen core approximation to reduce the active space.
This is particularly important in the analytic implementation because of the large
number of neutral excitations (poles) in $W_p$. This approximation has been shown to be very mild
in the past \cite{bruneval_jctc2013}.

To represent the 4-center Coulomb interaction,
we use the automatic generation of the auxiliary basis
designed in Ref.~\citenum{yang_jchemphys2007},
which corresponds to the ``PAUTO'' setting in Gaussian \cite{gaussian16}.

Besides this, there is no further approximation or technical convergence parameter in the analytic formula.
For the imaginary frequency quadrature, we follow the same recipes as those described hereafter for 
BAND.

\paragraph{BAND}
All calculations in BAND are performed with frozen core orbitals in all post-SCF parts of the calculation. The exceptions are the results reported in the SI\cite{supplmat} used to compare against FHI-AIMS, which are all-electron results. The $GW$ calculations are performed using the analytical frequency integration expression for the self-energy.\cite{tiago_prb2006,vansetten_jctc2013,bruneval_cpc2016} Only the quasiparticle self-consistent $GW$ qs$GW$\cite{Faleev2004, VanSchilfgaarde2006, Kotani2007} calculations are performed using the atomic-orbital based algorithm from Ref.~\citenum{Forster2021a}. 

All 4-center integrals are calculated using the pair-atomic density fitting scheme in the implementation of Ref.~\citenum{Spadetto2023}. The size of the auxiliary basis in this approach can be tuned by a single threshold which we set to $\varepsilon_{aux} = 1 \times 10^{-12}$ in all calculation if not stated otherwise. We further artificially enlarge the auxiliary basis by setting the BoostL option.\cite{Spadetto2023} 

In the implementation of the $G3W2$ contribution, we distinguish between the integration grid and the grid of values of $\mu + \icomp \omega$ at which the self-energy is evaluated using numerical integration. Both grids have been generated using the recipe in Ref.~\citenum{erhard_jchemphys2022}. 
To perform the integrations in a numerically stable way, it is necessary to split the expression according to the \emph{r.h.s.} of the second equation in Eq.~\eqref{eq:g3w2_split}. The double imaginary frequency integral only converges if it is restricted to the polarizable part of $W$. Therefore, we evaluate the contribution due to $GW_pGW_pG$, the polarizable part of SOSEX, $GW_pGvG$ and the static contribution $GvGvG$ separately using respectively Eqs.~\eqref{eq:g3w2_imag}, \eqref{eq:sosex_imag} and \eqref{eq:sox_imag} and calculate the full $G3W2$ self-energy according to Eq.~\eqref{eq:g3w2_split}.  
Even after splitting of the polarizable part of $W$, the double frequency integration converges slowly and we found that for some systems in the GW100 database up to 128 integration points are needed to converge this term. The actual number of frequency points used in each calculation is given in the supplemental information (Table~S2)
\cite{supplmat}.   

We evaluate the contributions to the $G3W2$ self-energy at 16 points along the $\mu + \icomp\omega$-axis and analytically continue them to the real axis. The SOSEX, 2SOSEX, and $G3W2$ self-energy corrections are then evaluated at the positions of the $GW$ QP energies. This corresponds to a linearization of the non-linear QP equations. This approximation is well justified in the valence region since the $GW$ QP peak will generally be very close to the $GW+G3W2$ peak and one expects the self-energy to be a relatively smooth function of $\omega$.\cite{foerster_prb2022}

\subsubsection{Comparison of MOLGW and BAND results}

In Table~\ref{tab:codes}, we report the HOMO energy of the Ne atom obtained from Hartree-Fock (HF) starting point
using the same basis set, namely Def2-TZVPP and 128 frequency points used in the numerical integrations. 
For all the self-energy approximations, $GW$, $GW$+SOSEX, or $GW+G3W2$, the agreement across the codes
and the numerical techniques is extremely good (1~meV at most).
The (tiny) differences between BAND and MOLGW can certainly be ascribed to the diverse auxiliary
bases used to expand the Coulomb interaction. 

As described in the SI,\cite{supplmat} the vertex corrections here are evaluated in a ``one-shot'' fashion at the position of the $GW$ QP energies. This approximation is implemented in BAND to make the calculations computationally feasible. In Table S4, we compare the BAND results to the analytical implementations in MOLGW with graphical solution of the QP equations. The results show that the effect of the one-shot approximation is small and that it can be used safely in all calculations.

Furthermore, with BAND, we have tested that the imaginary frequency formula in Eq.~(\ref{eq:g3w2_imag})
is not sensitive to the origin of the imaginary axis (provided it is inside the HOMO-LUMO gap).
This property is important for the sanity of the formula.
For instance, the vertex formula in Ref.~\citenum{wang_jctc2021} does not fulfill this 
property according to our evaluation.

Now that we have demonstrated the precision of our codes, we can turn the analysis of the
$G3W2$ properties in realistic molecular systems.

\begin{table}[b]
\caption{
HOMO energy of Ne within $GW+G3W2$ obtained with one-shot evaluation of the self-energies based on HF input.
A Def2-TZVPP basis set was used in both codes, MOLGW and BAND.
In BAND, we have tweaked the imaginary axis origin $\mu$ for stability check.
}
\label{tab:codes}
\begin{tabular}{lccccc}
\hline\hline
&
\multicolumn{2}{c}{MOLGW}
&
\multicolumn{3}{c}{BAND} \\
\cmidrule(lr){2-3}
\cmidrule(lr){4-6}
     & Analytic & Frequency grid &  Frequency grid & Frequency grid&  Frequency grid \\
     &          & $\mu$ = 0 eV     &   $\mu$ = -10 eV &  $\mu$ = 0 eV &  $\mu$ = +10 eV \\
$GW$ & -21.3513 & -21.3512       & -21.3504 & -21.3504 & -21.3504\\
$GW$+SOSEX
     & -21.9344 & -21.9349       & -21.9335 & -21.9335 & -21.9335 \\
$GW$+$G3W2$
     & -21.7214 & -21.7199       & -21.7200 & -21.7203 & -21.7204 \\
\hline\hline
\end{tabular}
\end{table}

\subsection{\label{sec:bench:basic}Basic properties}

\begin{figure}
 \includegraphics[width=0.99\columnwidth]{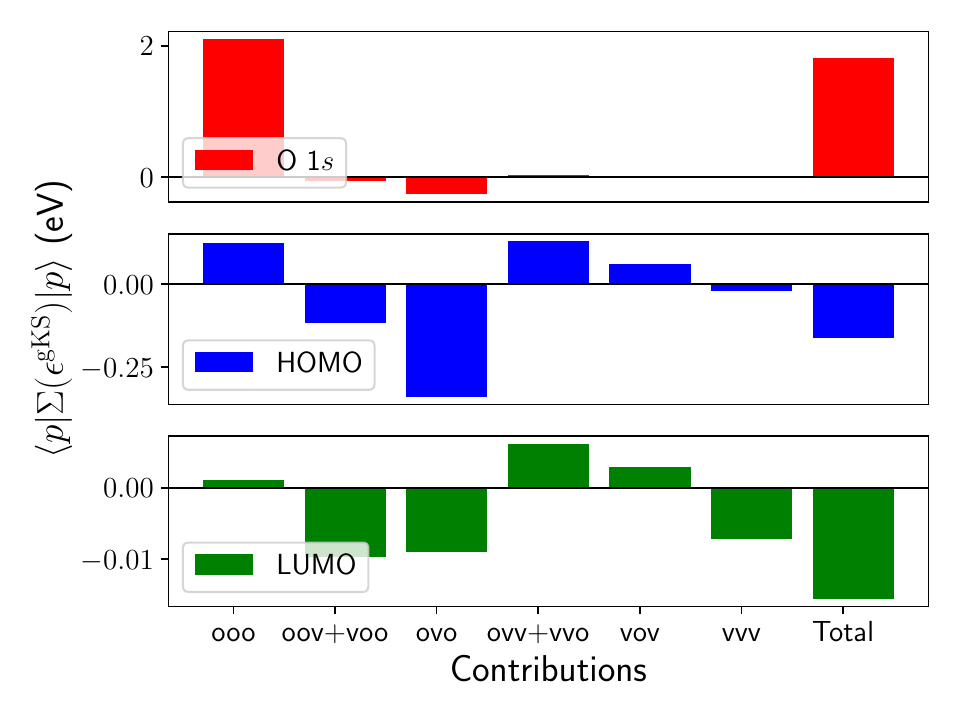}
  \caption{ 
      Decomposition of the $G3W2$ dynamical part into analytic terms for H$_2$O
      within pcSseg-3 basis based on PBEh(0.50).
      Expectation value for the O1$s$, the HOMO and the LUMO are given.
      Beware the change in y-axis scale.
    }
    \label{fig:contrib}
\end{figure}

Here we analyze the basic properties of the $G3W2$ self-energy for simple molecular systems.
First, we concentrate on the analytic formula in Eqs.~(\ref{eq:ooo}-\ref{eq:vov}).
While the computational scaling of all these terms is formally $N^7$, it is interesting to investigate
the magnitude of each of these terms.
In particular, in large basis sets, the number of virtual states $N_v$ is much larger than the number of
occupied states $N_o$ and we expect that the term $\Sigma^{vvv}$ in Eq.~\eqref{eq:vvv} will be the most expensive one to calculate.

In Fig.~\ref{fig:contrib}, we report the 6 terms in Eqs.~(\ref{eq:ooo}-\ref{eq:vov})
evaluated with PBEh(0.50) inputs for the water molecule (in the GW100 geometry \cite{vansetten_jctc2015}).
We show the expectation value of the self-energies at the corresponding PBEh(0.50) eigenvalue,
which is in general a very reasonable guess for valence states \cite{bruneval_cpc2016} and core states \cite{golze_jpcl2020}.
While the total $G3W2$ self-energy is large for a core state ($\sim$ 2 eV), it is rather weak
for the HOMO ($\sim$ 0.1 eV) and very weak for the LUMO ($\sim$ 0.01 eV).

For the core state, the term $\Sigma^{ooo}$ largely prevails.
$\Sigma^{vov}$, $\Sigma^{ovv}+\Sigma^{vvo}$, and $\Sigma^{vvv}$ are hardly visible on the scale of our plot.
They will be safely neglected in the core state evaluation in Section~\ref{sec:bench:core}.
Quite the opposite for valence electrons, the HOMO and LUMO states of H$_2$O
have similar weights for all the contributions and then none of them can be neglected.

\begin{figure}
 \includegraphics[width=0.49\columnwidth]{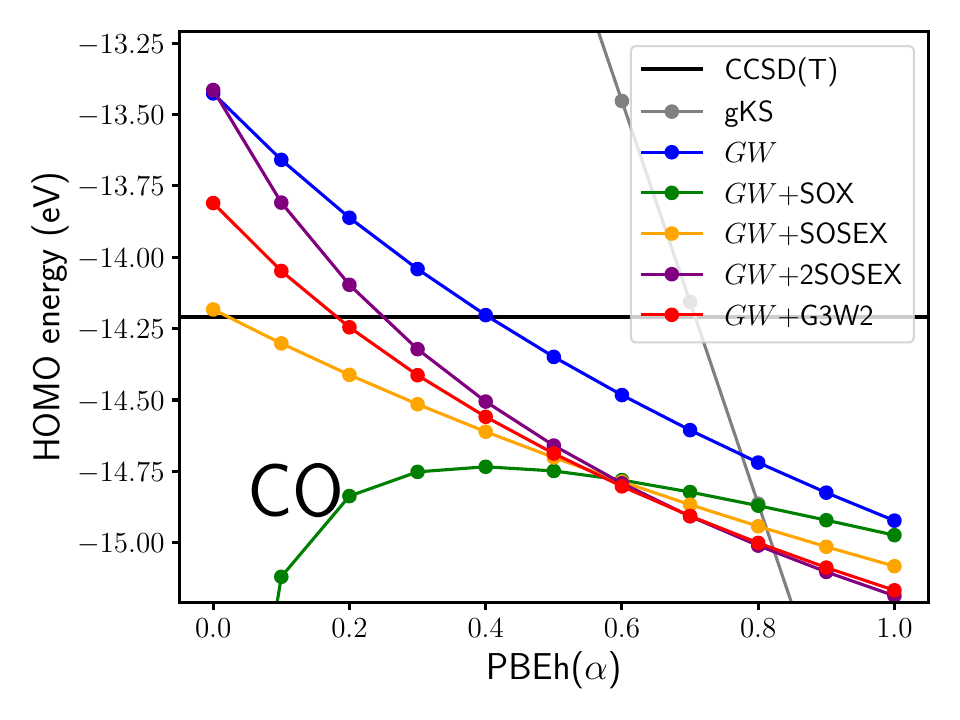}
 \includegraphics[width=0.49\columnwidth]{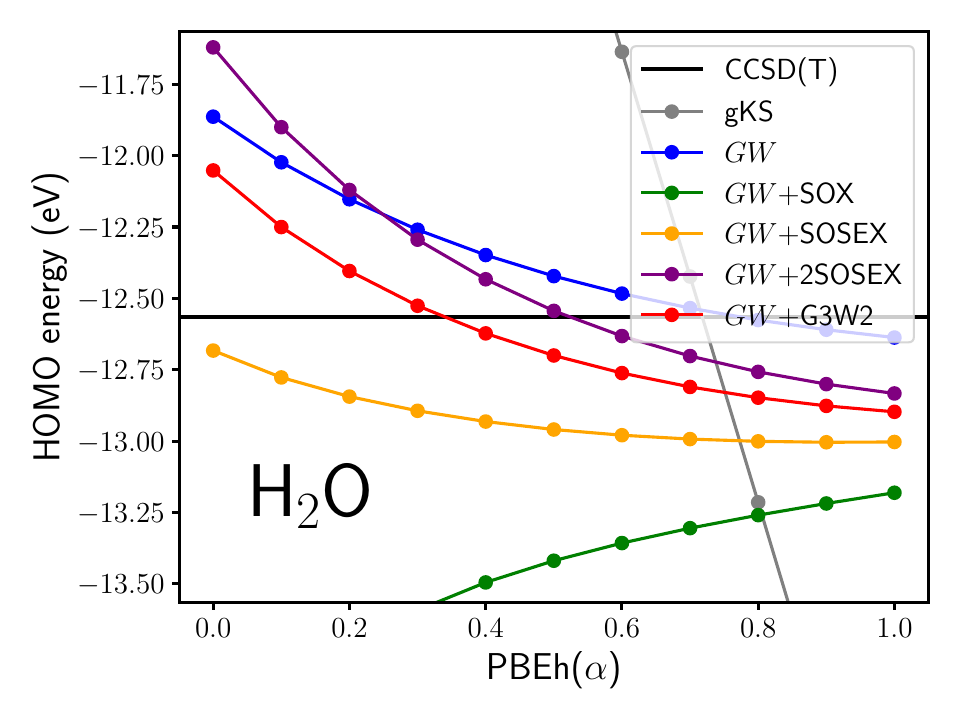}
  \caption{ 
    Starting point dependence of the HOMO quasiparticle energy for CO and H$_2$O
    with Def2-TZVPP basis set, for the different vertex approximation.
    The CCSD(T) reference \cite{bruneval_fchem2021} is shown as an horizontal line.
    }
    \label{fig:startingpoint}
\end{figure}

The starting point dependence of one-shot $GW$ (i.e. $G_0W_0$) is well documented in the literature
(see e.g. Ref.~\citenum{bruneval_jctc2013}).
The strong dependence of $GW$+SOSEX has also been studied by Ren and coworkers.\cite{ren_prb2013}.
In Fig.~\ref{fig:startingpoint} we analyze the starting point effect for all the vertex approximations
mentioned in this work $GW$ (no vertex), $GW$+SOX, $GW$+SOSEX, $GW$+2SOSEX, $GW+G3W2$ for the
HOMO energy of 2 molecules, CO and H$_2$O.
To do so, we tune the content of exact-exchange $\alpha$ in PBEh functional from 0 to 1,
with 0 giving the usual semi-local PBE functional and 0.25, PBE0.
For all the studied approximations, the effect of the starting point is large.
The PBE starting point, a terrible start for the quasiparticle energies of molecular systems,
quite luckily induces a $GW$+SOSEX HOMO that compares well with CCSD(T).
It is interesting to note that the $GW+G3W2$ curve is almost parallel to the $GW$ one:
the $G3W2$ induces then a constant downward shift.
However, for the starting points that yield good HOMO energies compared to
$GW$ or to CCSD(T) ($\alpha \sim $ 0.50 -- 0.75),
the $G3W2$ correction brings the result away from the correct value.
Only with a low value of $\alpha \sim 0.25$ can $GW+G3W2$ beat simple $GW$.

\subsection{$GW$100 ionization potentials}

\begin{figure}[hbt!]
    \centering
    \includegraphics[width=\columnwidth]{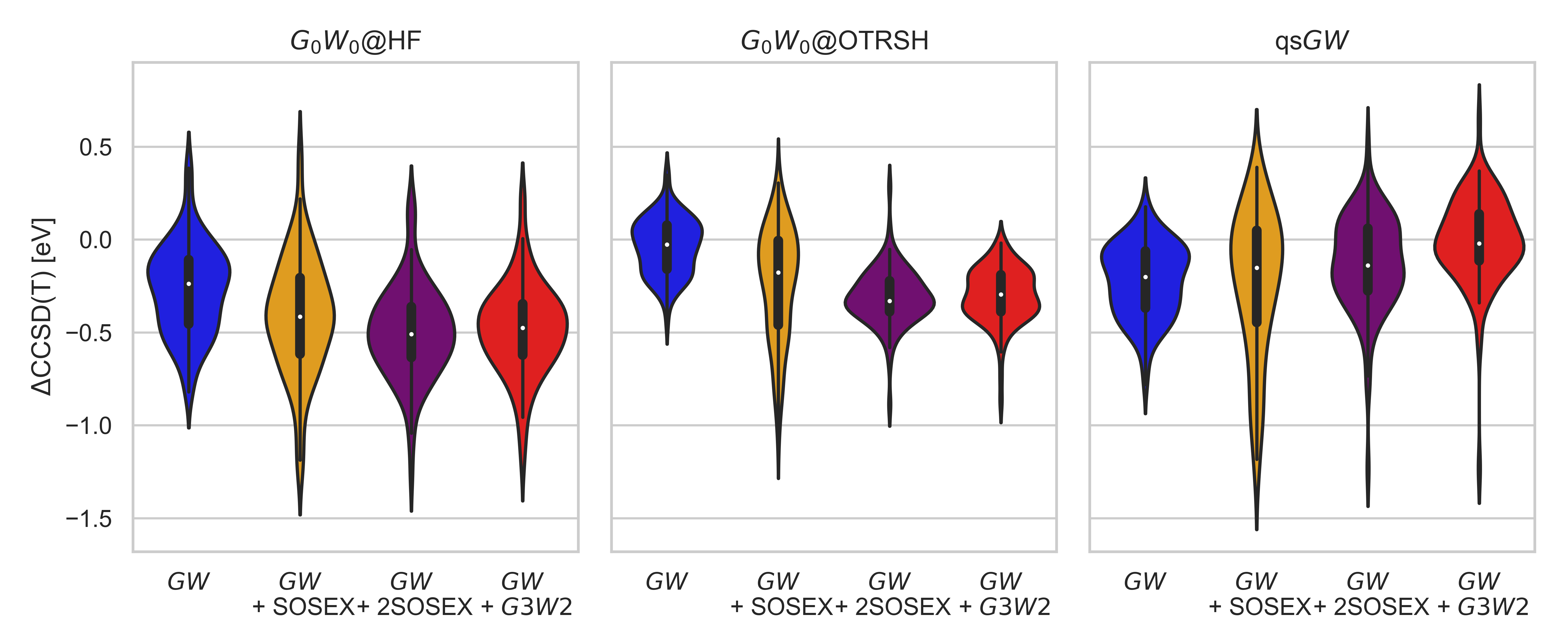}
    \caption{Errors of HOMO QP energies for $G_0W_0$ and different approximations of the $G3W2$ vertex correction with respect to CCSD(T) reference values \cite{bruneval_fchem2021} for GW100 using HF and OTRSH starting points. All values are in eV.}
    \label{fig:gw100_results}
\end{figure}

\begin{table}[hbt!]
    \centering
    \begin{tabular}{lcccc}
    \toprule
    & $GW$ 
    & $GW +$ SOSEX
    & $GW +$ 2SOSEX
    & $GW + G3W2$ \\
    \midrule
 HF    &  0.30 & 0.44 & 0.51 & 0.49 \\
 OTRSH &  0.13 & 0.28 & 0.32 & 0.30 \\
 qs$GW$&  0.23 & 0.32 & 0.21 & 0.17 \\ 
    \bottomrule
    \end{tabular}
    \caption{Mean absolute deviations of $G_0W_0$ and different approximations of the $G3W2$ vertex correction with respect to CCSD(T) reference values \cite{bruneval_fchem2021} for GW100 using HF and OTRSH starting points. All values are in eV.}
    \label{tab:gw100_results_MAD}
\end{table}

\subsubsection{Generalized Kohn-Sham starting points}

We focus first on the ``one-shot'' evaluation of the self-energy based on generalized Kohn-Sham.
There are an infinite variations of Kohn-Sham starting points that could be used.
Here we decide to concentrate on two remarkable ones: Hartree-Fock (HF) and
optimally-tuned range-separated hybrid (OTRSH) \cite{baer_prl2005,livshits_pccp2007}.
Firstly, HF is the most commonly used starting point for ``post Hartree-Fock'' methods in the quantum chemistry. Though not the most accurate choice, this approximation is useful for comparison and
benchmarking.
Secondly, OTRSH is less used in the literature however it has been shown recently to minimize the mean-error
for $G_0W_0$ IPs for the GW100 set \cite{McKeon2022}.
In the present work we use a PBE-based version of OTRSH \cite{refaely_prl2012,kronik_jctc2012}
where the short-range exchange is that of Henderson and coworkers \cite{henderson_jchemphys2008}.
The range-separation parameter $\gamma$ is determined so that the ``$GW$ correction'' vanishes
as proposed in Ref.~\citenum{McKeon2022},
while exact-exchange scaling parameters $\alpha=0.2$ and $\beta=0.8$ are kept constant.

Figure~\ref{fig:gw100_results} shows the errors distributions of the $G_0W_0$, $G_0W_0$~+~SOSEX, $G_0W_0$~+~2SOSEX and $G_0W_0 + G3W2$ HOMO QP energies for HF and OTRSH starting points compared to the CCSD(T) reference values of Ref.~\citenum{bruneval_fchem2021} for the GW100 set.
Mean absolute deviations (MAD) for all methods are shown in Table~\ref{tab:gw100_results_MAD}. With a MAD of 0.13 eV, $G_0W_0$@OTRSH is among the best performing $GW$ methods for the GW100 set\cite{McKeon2022} and any vertex corrections to the self-energy worsens the agreement with CCSD(T). The same observation can be made for $G_0W_0$@HF. 
For both starting points, the addition of SOSEX leads to higher IPs (more negative HOMOs) on average and also the spread of errors increases substantially, as can be seen from the distributions in Figure~\ref{fig:gw100_results}. Symmetric screening of the SOX term in the form of $G_0W_0$~+~2SOSEX slightly increases the average error compared to $G_0W_0$+SOSEX. However, it reduces the spread of the error again, indicating the 2SOSEX is a more consistent approximation to the full $G3W2$ term than SOSEX alone. 
Finally, the addition of the double frequency integral, leading to the full $G_0W_0 + G3W2$ term, has only little effect on the MADs and the shape of the error distributions. This result has already been suggested by Figure~\ref{fig:startingpoint} which shows that for large content of exchange $\alpha$, 2SOSEX and $G3W2$ give similar results. Also Table~S2 \cite{supplmat} shows that the contribution from the double frequency integral is typically of the order of only a few tens meV and only exceeds 100 meV in very few cases.

This observation suggests that 2SOSEX would be a consistent and suitable approximation to the complete $G3W2$ term. However, our data also clearly shows that it does not improve over $G_0W_0$.

\subsubsection{Quasiparticle self-consistent $GW$ starting point}

Turning to the qs$GW$ results shown in Fig.~\ref{fig:gw100_results} we observe that the vertex correction shifts the qs$GW$ results in the opposite direction than for $G_0W_0$. qs$GW$ alone overestimates the CCSD(T) IPs, but qs$GW$ + $G3W2$ lowers them, bringing them in better agreement with the reference values. However, the vertex correction also produces clear outliers. One of them is Helium, one of the simplest systems in GW100, for which qs$GW$ + $G3W2$ deviates from CCSD(T) by more than 1.2 eV. This shows that $G3W2$ is also not a robust method to improve over qs$GW$. 
Our construction of the qs$GW$ Hamiltonian is based on ref.~\citenum{Kotani2007}, even though alternative constructions have been shown to give slightly improved results for GW100.\cite{Marie2023a} 

In Table~S5\cite{supplmat} we also present results for qsGW$_0$ for a few selected systems, where $W$ is kept fixed after the first SCF cycle. These results show that the update of $W$, not the update of $G$ is responsible for the qualitatively different behavior of qs$GW$ compared to $G_0W_0$. As for $G_0W_0$, the $G3W2$ diagram increases the qsGW$_0$ IPs.

Finally, we notice that vertex corrections to qs$GW$ for periodic semi-conductors and insulators have so far only been investigated in the polarizability\cite{Cunningham2018, Cunningham2021, Tal2021, Lorin2023}. This is due to a well-known argument\cite{Kotani2007} that a non-interacting Green's function behaves as an effective vertex correction to the self-energy. Therefore, the vertex should not be added to the self-energy to avoid that it is effectively double counted. Our results question the validity of this argument.

\subsection{ACC24 ionization potentials and electron affinities}

After having analyzed the $G3W2$ term for GW100, we turn to the ACC24 set\cite{richard_jctc2016} of 24 organic acceptor molecules with stable anions. This also allows for a meaningful comparison of LUMO QP energies and HOMO-LUMO gaps. We focus here on $G_0W_0$ based on the OTRSH starting point which has been shown to give very good results also for the ACC24 set.\cite{knight_jctc2016}

\begin{figure}[hbt!]
    \centering
    \includegraphics[width=\columnwidth]{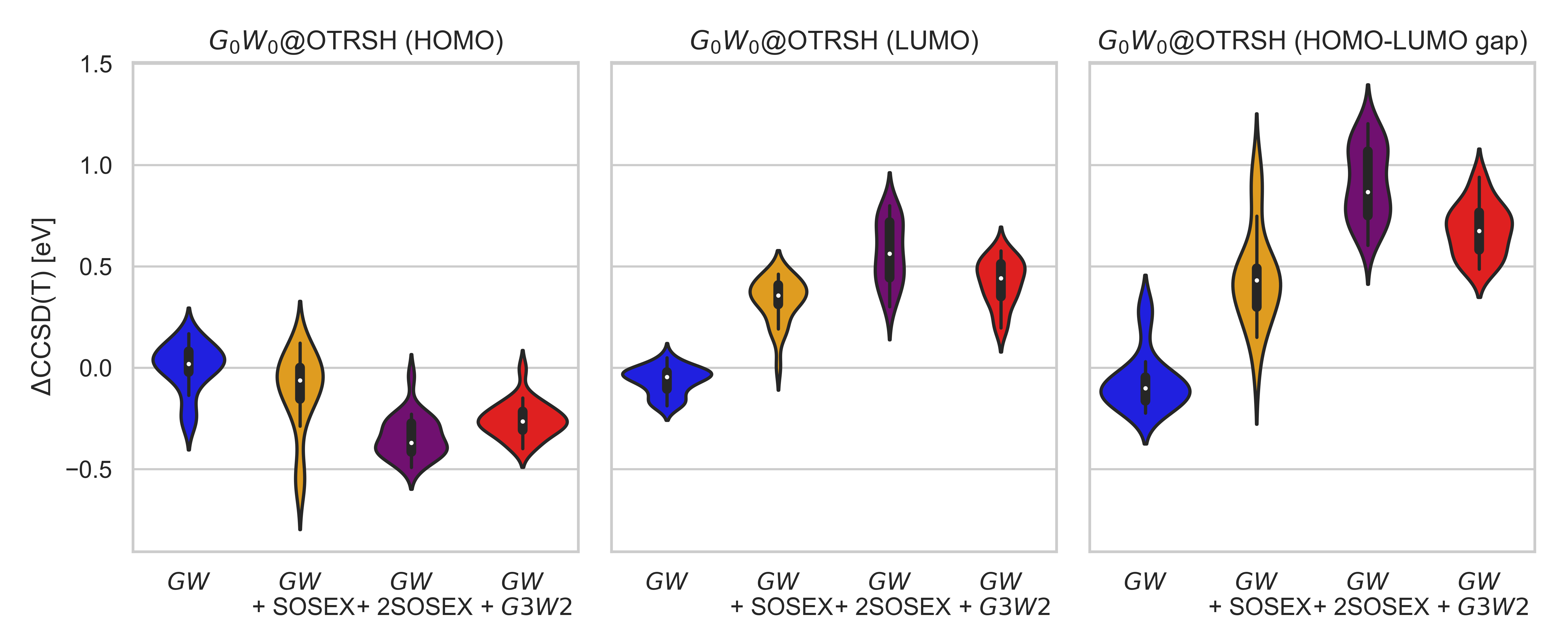}
    \caption{Errors of $G_0W_0$@OTRSH with different approximations of the $G3W2$ vertex correction with respect to CCSD(T) reference values for ACC24 for HOMO, LUMO and HOMO-LUMO gaps. All values are in eV. For some molecules, the $G3W2$ corrections have been performed using cc-pVTZ or aug-cc-pVDZ due to prohibitively high costs. See supporting information for details.}
    \label{fig:acc24_results}
\end{figure}

\begin{table}[hbt!]
    \centering
    \begin{tabular}{lcccc}
    \toprule
    & $GW$ 
    & $GW +$ SOSEX
    & $GW +$ 2SOSEX
    & $GW + G3W2$ \\
    \midrule
HOMO          &  0.09 & 0.15 & 0.34 & 0.26 \\
LUMO          &  0.07 & 0.33 & 0.57 & 0.42 \\
HOMO-LUMO gap &  0.14 & 0.46 & 0.91 & 0.68 \\ 
    \bottomrule
    \end{tabular}
    \caption{Mean absolute deviations of $G_0W_0$@OTRSH and different approximations of the $G3W2$ vertex correction with respect to CCSD(T) reference values for ACC24 for HOMO, LUMO, and HOMO-LUMO gap. All values are in eV.}
    \label{tab:acc24_results_MAD}
\end{table}

For the OTRSH starting point, Figure~\ref{fig:gw100_results} shows the errors distributions of the $G_0W_0$, $G_0W_0$~+~SOSEX, $G_0W_0$~+~2SOSEX and $G_0W_0 + G3W2$ HOMO, LUMO and HOMO-LUMO gaps using against the CCSD(T) reference values of Ref.~\citenum{richard_jctc2016} for the ACC24 set. The corresponding MADs are shown in Table~\ref{tab:acc24_results_MAD}. We do not use the basis set limit extrapolated reference values from Ref.~\citenum{richard_jctc2016} but instead CCSD(T)/aug-cc-pVTZ results as reference. For the four molecules for which results with aug-cc-pVTZ are not available, we compare against CCSD(T)/aug-cc-pVDZ. We notice that also a comparison at the aug-cc-pVTZ does not eliminate basis set errors since the CCSD(T) converges faster to the complete basis set limit than $GW$ for individual QP energies.\cite{bruneval_jctc2013}
Only for HOMO-LUMO gaps, the rates of convergence are comparable. 

Also, due to the high computational cost we calculate the vertex corrections at aug-cc-pVTZ for only about half of the 24 molecules while for the other we use either cc-pVTZ or aug-cc-pVDZ. For a discussion of the basis set limit convergence and details on the choice of basis for each system we refer to Tables S7 and S8.
\cite{supplmat}
For the molecules for which we can afford aug-cc-pVTZ, we see that the results are almost identical to cc-pVTZ (MADs of 5-6 meV for HOMO and LUMO for the different vertex corrections). Also aug-cc-pVDZ is relatively close to TZ, with MADs of 60 meV for HOMO and 90 meV for LUMO for the full $G3W2$ correction. We therefore believe that this comparison reveals the qualitative behavior of the vertex corrections we test here. 

For the HOMO energies we observe the same trend as for the GW100 set and notice that $G_0W_0$@OTRSH reproduces the CCSD(T) reference values with a MAD of 0.09 eV very well. The same can be observed for the LUMO energies. Here, the vertex corrections increase the LUMO energies considerably. In combination with the decreased HOMO energies, this leads to a major increase of the HOMO-LUMO gaps. The $G_0W_0$@OTRSH $+ G3W2$ HOMO-LUMO gap has an average signed error of 0.68 eV, more than 0.5 eV worse than $G_0W_0$@OTRSH with 0.14 eV only. 

\subsection{\label{sec:bench:core}Core levels}

\begin{figure}
 \includegraphics[width=0.49\columnwidth]{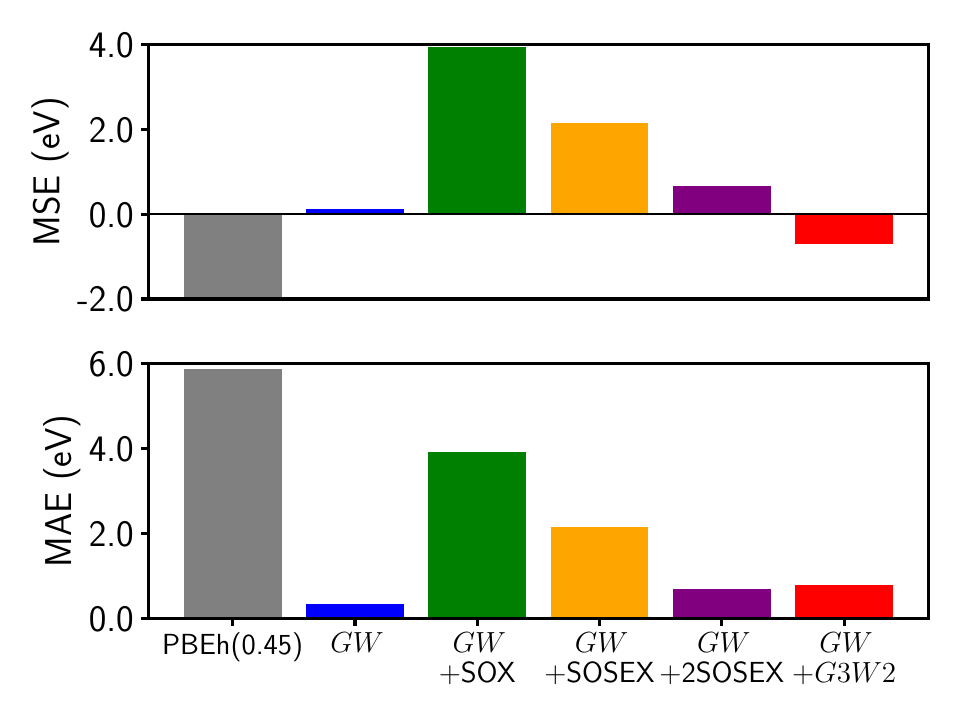}
  \caption{Mean signed deviation (upper panel) and mean absolute deviation (lower panel) in eV
  for 28 core state energies taken among the smallest molecules in CORE65 benchmark set
      \cite{golze_jpcl2020} evaluated within pc-Sseg-3 basis based on PBEh(0.45).
      This plot uses binding energies, which is the negative of the quasiparticle energies, as opposed
      to the rest of this study.
      Note that the PBEh(0.45) bars in gray are truncated for the sake of visualization.
    }
    \label{fig:core65}
\end{figure}

Core level binding energies have been recently addressed in the framework of 
$GW$ \cite{golze_jctc2018,vansetten_jctc2018,golze_jpcl2020}.
We use the opportunity offered by the fully analytic expressions reported in Eqs.~(\ref{eq:ooo})-(\ref{eq:vov})
to accurately evaluate $GW$ and $G3W2$ self-energies for core levels.

Building on the work of Meija-Rodriguez \textit{et. al.} \cite{mejia_jctc2022},
we employ an accurate basis set that was designed in particular to describe core electrons,
namely pc-Sseg-3.
Using Golze's recommendation \cite{golze_jpcl2020}, we evaluate the 
self-energy using a one-shot procedure based on PBEh(0.45).
Relativistic corrections are added \textit{a posteriori} with the values prescribed by
Golze and coworkers \cite{golze_jpcl2020}.
In the CORE65 set \cite{golze_fchem2019}, the reference binding energies are taken from various experiments.

The calculations based on the analytic approach to $G3W2$ are very demanding
and hereafter we neglect the contributions to the self-energy that involve two or three virtual orbitals.
We have seen in Sec.~\ref{sec:bench:basic} that these terms are of the order of a few meV for core levels.
Even with this approximation, the calculations remain challenging.
Figure~\ref{fig:core65} reports
the MSD and MAD for 28 core states out of the 65 energies gathered in the original CORE65 set.
We have excluded the largest molecules because of computational burden and O$_2$ because of spin.
The comprehensive results are given in Table S9 in the supplemental material\cite{supplmat}.

From Figure~\ref{fig:acc24_results}, we see that $GW$ yields the best result
and that $GW$+SOX deteriorates much.
Gradually adding screening through SOSEX, 2SOSEX, and then full $G3W2$ goes in the correct direction.
For core levels, the vertex corrections are indeed sizeable.
It should be noted  though, that the PBEh(0.45) starting point was specifically selected
in Ref.~\citenum{golze_jpcl2020} to minimize the $G_0W_0$ error.
Tuning $\alpha$ in PBEh($\alpha$) could improve $GW+G3W2$.
But such a study is beyond the scope of the present work.

%%%%%%%%%%%%%%%%%%%%%%%%%%%%%%%%%%%%%%%%%%%%%%%%%%%%%%%%%%%%%%%%%%%%%%%%%%
\section{Conclusion}
\label{sec:conclusion}
%%%%%%%%%%%%%%%%%%%%%%%%%%%%%%%%%%%%%%%%%%%%%%%%%%%%%%%%%%%%%%%%%%%%%%%%%%%

In this work, we have derived and evaluated the formulas for the self-energy diagram
drawn in Figure~\ref{fig:diag_g3w2},
which contains
two dynamically screened Coulomb interactions $W(\omega)$ and three Green's functions:
this is the diagram named $G3W2$ by Kutepov \cite{kutepov_jpcm2021}.
We propose two formulas: a fully analytic one that is useful for analysis, reference and core levels
and a numerical one that uses a double complex frequency quadrature that is more efficient for the HOMO-LUMO region.
% The latter expression departs from the single complex frequency quadrature published earlier by Wang \textit{et. al.} \cite{wang_jctc2021}.

Based on well-established molecular benchmarks for ionization potentials (GW100)\cite{vansetten_jctc2015},
electron affinities (ACC24)\cite{knight_jctc2016}, and core levels (CORE65)\cite{golze_jpcl2020},
we conclude that in general the $GW+G3W2$ self-energy does not improve over the simpler $GW$ self-energy.
When designing simpler semi-static approximations to $G3W2$,
in the spirit of SOSEX \cite{ren_prb2015},
we propose a new approximation that we name 2SOSEX that better approximates the complete $G3W2$ self-energy.

The strategy to keep the RPA screening and introduce vertex corrections only in the self-energy
is certainly not sufficient. As already suggested long time ago by Del Sole and coworkers, vertex corrections in the self-energy should be combined with vertex corrections in $W$.\cite{delsole_prb1994}.
This should be the subject of future investigation.

\begin{acknowledgement}
We acknowledge constructive discussions with Xinguo Ren and Patrick Rinke, aimed at clarifying that their equation (11) \cite{wang_jctc2021} does not represent the exact expression corresponding to the complete second-order diagram.
Part of this work was performed using HPC resources from GENCI–TGCC (Grant 2023-gen6018). 
AF acknowledges the use of supercomputer facilities at SURFsara sponsored by NWO Physical Sciences, with financial support from The Netherlands Organization for Scientific Research (NWO).
\end{acknowledgement}

\begin{suppinfo}
Individual results for each molecule in the GW100, ACC24, CORE65 benchmarks are made as Supporting information.
Frequency grid convergence and comparison to FHI-AIMS results are also shown.
\end{suppinfo}

\appendix

%%%%%%%%%%%%%%%%%%%%%%%%%%%%%%%%%%%%%%%%%%%%%%%%%%%%%%%%%%%%%%%%%%%%%%%%%%
\section{Fourier-transforming $\Sigma^{G3W{_p}2}$ to real frequencies}
\label{app:realfreqs}
%%%%%%%%%%%%%%%%%%%%%%%%%%%%%%%%%%%%%%%%%%%%%%%%%%%%%%%%%%%%%%%%%%%%%%%%%%%

This appendix shows all the steps to obtain the frequency-dependent expression of the
dynamic part of the $G3W2$ self-energy, starting from the expression with times as
obtained from Hedin's equation.

Let us recall here Eq.~(\ref{eq:g3w2_times}):
\begin{equation}
 \Sigma^{G3{W_p}2}(t_1-t_2)
    = \icomp^2 \int dt_3 dt_4 G(t_1 -t_4) W_p(t_3-t_1) G(t_4-t_3) W_p(t_2-t_4) G(t_3-t_2)  
\end{equation}
and let us introduce the difference time variables:
\begin{subequations}
    \begin{eqnarray}
         \tau &=& t_1 - t_2 \\
  \tau^\prime &=& t_1 - t_3 \\
\tau^{\prime\prime} &=& t_4- t_2
\end{eqnarray}

Combining these 3 variables, we obtain the other 3 needed time differences:
\begin{eqnarray}
  t_1 - t_4 &=&  \tau - \tau^{\prime\prime} \\
  t_4 - t_3 &=& -\tau + \tau^\prime + \tau^{\prime\prime}\\
  t_3 - t_2 &=&  \tau - \tau^\prime .
\end{eqnarray}
\end{subequations}

%Change of integration variable from $t_3, t_4$ to $\tau^\prime, \tau^{\prime\prime}$:
%\begin{subequations}
%\begin{eqnarray}
% \int_{-\infty}^{+\infty} d t_3 & = &  -\int_{+\infty}^{-\infty} d \tau^\prime
%   =  \int_{-\infty}^{+\infty} d \tau^\prime \\
% \int_{-\infty}^{+\infty} d t_4 & = &  \int_{-\infty}^{+\infty} d \tau^{\prime\prime} 
%\end{eqnarray}
%\end{subequations}

Integration over $t_3$ and $t_4$ are then transformed to $\tau^\prime$ and $\tau^{\prime\prime}$
without other change:
\begin{equation}
\label{eq:sigmaGWGWG_tau}
 \Sigma(\tau) = \icomp^2 \int d \tau^\prime d \tau^{\prime\prime}
     G(\tau - \tau^{\prime\prime}) W_p(-\tau^\prime)
     G(-\tau + \tau^\prime + \tau^{\prime\prime})
      W_p(-\tau^{\prime\prime})
      G(\tau - \tau^\prime)  .
\end{equation}

At the stage, we Fourier-transform all the times using the following definitions:
\begin{subequations}
\begin{equation}
  G(\omega) = \int d \tau e^{\icomp \omega \tau} G(\tau) 
\end{equation}
and
\begin{equation}
  G(\tau) = \frac{1}{2\pi}  \int d \omega e^{-\icomp \omega \tau} G(\omega)  .
\end{equation}
\end{subequations}

$\Sigma$ is forward-transformed whereas the $G$ and the $W_p$ are backward-transformed:
\begin{multline}
\Sigma(\omega) = \frac{\icomp^2}{(2\pi)^5} \int d \tau d \tau^\prime d\tau^{\prime\prime} 
     e^{\icomp \omega \tau}
\int d \omega_1 d \omega^{\prime} d\omega_2 d\omega^{\prime\prime} d\omega_3 \\
\times
     e^{-\icomp \omega_1 (\tau-\tau^{\prime\prime})}
     e^{+\icomp \omega^\prime \tau^{\prime}}
     e^{-\icomp \omega_2 (-\tau+\tau^{\prime}+\tau^{\prime\prime})}
     e^{+\icomp \omega^{\prime\prime} \tau^{\prime\prime}}
     e^{-\icomp \omega_3 (\tau-\tau^{\prime})}   \\ 
    \times  G(\omega_1) W_p(\omega^\prime) G(\omega_2) W_p(\omega^{\prime\prime})  G(\omega_3)
\end{multline}

Performing the integrals over $\tau, \tau^\prime, \tau^{\prime\prime}$ variables yields 3 $\delta$ functions:
\begin{subequations}
\begin{eqnarray}
 \int d \tau e^{\icomp \tau (\omega - \omega_1 + \omega_2 - \omega_3 ) }
     & = & 2\pi\delta(\omega - \omega_1 + \omega_2 - \omega_3 ) \\
 \int d \tau^\prime e^{\icomp \tau^\prime (\omega^\prime - \omega_2 + \omega_3 ) }
     & = & 2\pi\delta(\omega^\prime - \omega_2 + \omega_3 ) \\
      \int d \tau^{\prime\prime} e^{\icomp \tau^{\prime\prime} (\omega^{\prime\prime} + \omega_1 - \omega_2 ) }
     & = & 2\pi\delta(\omega^{\prime\prime} + \omega_1 - \omega_2) .
\end{eqnarray}
\end{subequations}

This allows one to eliminate $\omega_1$, $\omega_2$, and $\omega_3$:
\begin{subequations}
\begin{eqnarray}
 \omega_1 &=& \omega + \omega^\prime \\ 
 \omega_2 &=& \omega + \omega^\prime + \omega^{\prime\prime}\\
 \omega_3 &=& \omega \phantom{ + \omega^\prime} ~\, + \omega^{\prime\prime}  .
\end{eqnarray}
\end{subequations}

Finally, this yield the result announced in the main text in Eq.~(\ref{eq:g3wp2}):
\begin{equation}
  \Sigma^{G3W{_p}2}(\omega) =
    \left( \frac{\icomp}{2\pi} \right)^2
     \int d\omega^\prime \int d\omega^{\prime\prime}
        G(\omega+\omega^\prime)
        W_p(\omega^\prime) 
        G(\omega+\omega^\prime+\omega^{\prime\prime})
        W_p(\omega^{\prime\prime})
        G(\omega+\omega^{\prime\prime}) .
\end{equation}

%%%%%%%%%%%%%%%%%%%%%%%%%%%%%%%%%%%%%%%%%%%%%%%%%%%%%%%%%%%%%%%%%%%%%%%%%%
\section{Analytic calculation of $\Sigma^{ooo}$}
\label{app:ooo}
%%%%%%%%%%%%%%%%%%%%%%%%%%%%%%%%%%%%%%%%%%%%%%%%%%%%%%%%%%%%%%%%%%%%%%%%%%%

\begin{figure}
  \includegraphics[width=0.40\columnwidth]{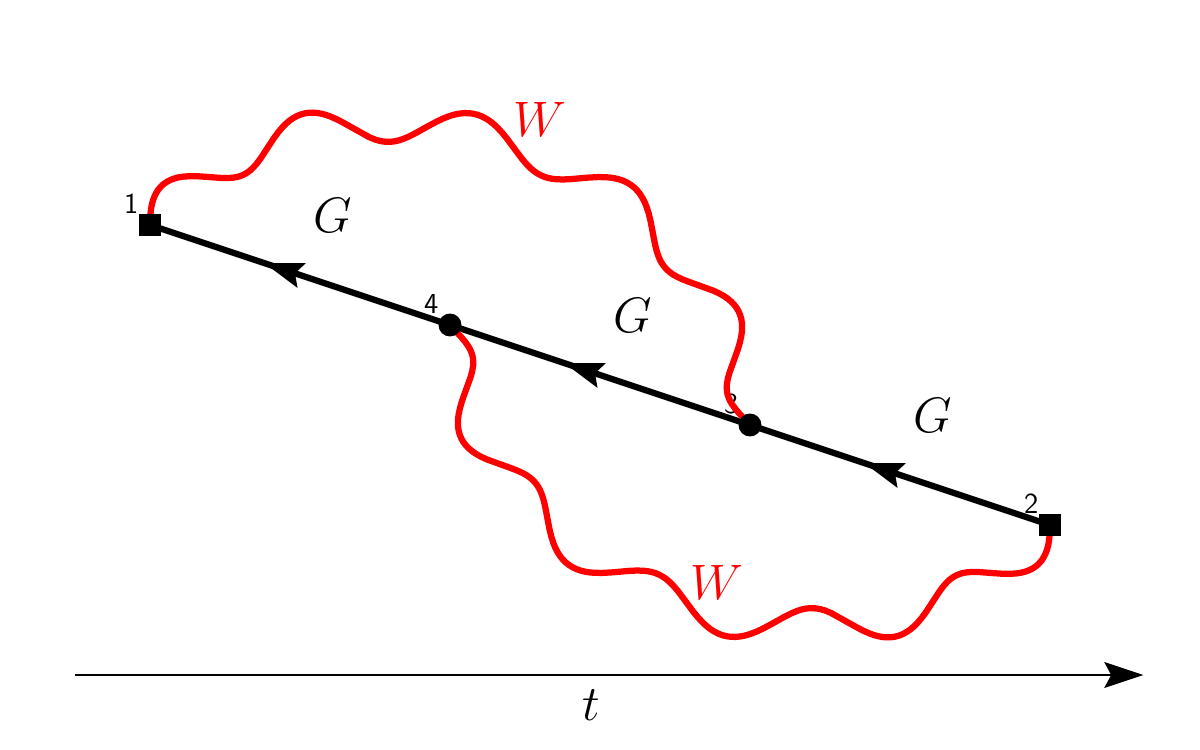}
  \caption{Goldstone-Feynman diagram for $\Sigma^{ooo}$.}
  \label{fig:diag_ooo}
\end{figure}

In this appendix, we write the step-by-step derivation of the part of the self-energy that involves
3 occupied MO in $G$, labeled $i$, $j$, $k$ and 2 resonant poles in $W_p$, labeled $t$ and $s$.
The corresponding Feynman-Goldstone diagram is drawn in Figure~\ref{fig:diag_ooo}, where
 the time arrow is shown.
 The analytic expression of this diagram involve the following frequency parts:
\begin{multline}
\label{eq:res0}
 \frac{\icomp}{2\pi} \int d\omega^\prime  
       \frac{\icomp}{2\pi} \int d\omega^{\prime\prime} 
       \frac{1}{\omega+\omega^\prime-\epsilon_i - \icomp \eta}
    \times    \frac{1}{\omega^\prime - \Omega_t + \icomp \eta}  \\
   \times  \frac{1}{\omega+\omega^\prime+\omega^{\prime\prime}-\epsilon_j - \icomp \eta} 
    \times   \frac{1}{\omega^{\prime\prime} - \Omega_s + \icomp \eta}  
     \times  \frac{1}{\omega+\omega^{\prime\prime}-\epsilon_k - \icomp \eta}
\end{multline}

First we perform the integral $\frac{\icomp}{2\pi} \int d\omega^{\prime\prime}$ using the residue theorem.
We draw a contour consisting the real axis and an arc below the real axis.
This contour encloses the poles with negative imaginary part and there is only one:
$\omega^{\prime\prime} =  \Omega_s - \icomp \eta$.
The residue theorem then gives
\begin{equation}
\label{eq:res1}
  -2\pi \icomp
   \frac{1}{\omega+\omega^\prime + \Omega_s -\epsilon_j - 2\icomp \eta} \times
   \frac{1}{\omega + \Omega_s -\epsilon_k - 2\icomp \eta} .
\end{equation}
The leading negative sign comes from the anticlockwise closing of the contour.
The second term in Eq.~(\ref{eq:res1}) has no $\omega^\prime$ dependence and is a simple prefactor.

Second we have to perform the integral over $\omega^\prime$:
\begin{equation}
 \frac{\icomp}{2\pi} \int d\omega^\prime
 \frac{1}{\omega+\omega^\prime-\epsilon_i - \icomp \eta} \times
 \frac{1}{\omega^\prime - \Omega_t + \icomp \eta} \times
 \frac{1}{\omega+\omega^\prime + \Omega_s -\epsilon_j - 2\icomp \eta}.
\end{equation}
The first two factors come from Eq.~(\ref{eq:res0}) and the last one from Eq.~(\ref{eq:res1}).

We use the residue theorem again with a contour consisting of the real axis and an arc in the lower part of the complex plane. The contour is oriented anticlockwise.
There is only one pole with a negative imaginary part:
$\omega^{\prime} =  \Omega_t - \icomp \eta$.
The application of the residue theorem reads
\begin{equation}
\label{eq:res2}
  -2\pi \icomp
   \frac{1}{\omega + \Omega_t -\epsilon_i - 2 \icomp \eta} \times
   \frac{1}{\omega + \Omega_t + \Omega_s -\epsilon_j - 3\icomp \eta} .
\end{equation}

Gathering the two factors in Eq.~(\ref{eq:res2}) and the last factor in Eq.~(\ref{eq:res1}),
we obtain the complete frequency dependence:
\begin{equation}
  \frac{1}{\omega-\epsilon_i+\Omega_t - 2\icomp \eta}
  \times
  \frac{1}{\omega-\epsilon_j+\Omega_t+\Omega_s - 3\icomp \eta}
  \times
  \frac{1}{\omega-\epsilon_k+\Omega_s - 2\icomp \eta} .
\end{equation}

Now inserting the correct spatial numerators, we get the final expression for 
$\Sigma_{pq}^{ooo}(\omega)$:
\begin{equation}
\Sigma_{pq}^{ooo}(\omega)
=
\sum_{t s}
\sum_{i j k } 
\frac{w_t^{pi} \cdot w_t^{jk} \cdot w_s^{qk} \cdot w_s^{ij}}
{(\omega  - \epsilon_i + \Omega_t - 2 \icomp \eta) \cdot (\omega - \epsilon_j + \Omega_t + \Omega_s  - 3 \icomp \eta )
\cdot (\omega - \epsilon_k + \Omega_s  - 2\icomp \eta)} .
\end{equation}

This is the complete occupied-occupied-occupied self-energy.
Indeed, the other terms that involve anti-resonant excitations in $W_p$
have all the poles located in the same upper-half of the complex plan and therefore vanish.

\section{Fourier-transforming $\Sigma^{G3W{_p}2}$ to imaginary frequencies}
\label{app:imagfrequencies}
The derivation of \eqref{eq:g3w2_imag} is very similar to the derivation of \eqref{eq:g3wp2} described in detail in appendix~\ref{app:realfreqs}. The only difference is than one needs to work with Laplace transforms which connect the real time axis to the complex plane. While these expressions are generally not unique, we choose a definition which matches the expression often used in space-time formulations of $GW$\cite{rieger_cpc1998}
\begin{subequations}
\label{laplace-transform}
    \begin{equation}
        G(\mu + \icomp \omega) = \icomp \int^{\infty}_{-\infty} e^{-(\mu + \icomp \omega)\tau} G(\tau) d \tau
    \end{equation}
\label{laplace-transform-back}
    \begin{equation}
        G(\tau) = -\frac{\icomp}{2\pi} 
        \int^{\infty}_{-\infty} 
        e^{(\mu + \icomp \omega)\tau} G(\mu + \icomp \omega) d \omega \;.
    \end{equation}
\end{subequations}
$\tau$ is a real time argument. Both transformations are well-defined as long as $\mu$ separates the HOMO and LUMO energies. 
Consistent with these definitions, the imaginary-frequency single-particle Green's function can be defined by 
\begin{equation}
\label{eq:g_complex}
    G(\br,\br', \mu + \icomp \omega) = \sum_{p} \frac{\phi_p(\br) \phi^*_p(\br')}{\mu + \icomp \omega - \epsilon_p} \;.
\end{equation}
To arrive at \eqref{eq:g3w2_imag}, one starts from \eqref{eq:sigmaGWGWG_tau} and forward transforms $\Sigma$ using \eqref{laplace-transform}, while all $G$ and $W_p$ are backward-transformed. One then arrives at (omitting spatial indices for brevity) 
\begin{equation}
\begin{aligned}
\Sigma^{G3W2}(\mu + \icomp\omega) = 
    \left(\frac{-1}{2\pi}\right)^2 
    \int d\omega^{\prime} d\omega^{\prime\prime} & 
    G(\mu + \icomp\omega + \icomp\omega^{\prime})
    G(\mu + \icomp\omega + \icomp\omega^{\prime} + \icomp\omega^{\prime\prime})\\ 
    & \times 
    G(\mu + \icomp\omega + \icomp\omega^{\prime\prime}) 
    W(\icomp\omega^{\prime})
    W(\icomp\omega^{\prime\prime}) \;,
\end{aligned} 
\end{equation}
and using \eqref{eq:g_complex} for all $G$ one arrives at \eqref{eq:g3w2_imag}.

%\bibliography{my_biblio, arno_ref}
\providecommand{\latin}[1]{#1}
\makeatletter
\providecommand{\doi}
  {\begingroup\let\do\@makeother\dospecials
  \catcode`\{=1 \catcode`\}=2 \doi@aux}
\providecommand{\doi@aux}[1]{\endgroup\texttt{#1}}
\makeatother
\providecommand*\mcitethebibliography{\thebibliography}
\csname @ifundefined\endcsname{endmcitethebibliography}
  {\let\endmcitethebibliography\endthebibliography}{}

\end{document}